\documentclass[a4paper,11pt]{article}
\usepackage{jheppub, bm, color} 
\usepackage{amssymb,amsfonts,slashed,amsthm,amsmath,graphicx, soul}
\usepackage[caption=false]{subfig}
\newcommand{\MeV}{\  {\rm MeV} }
\newcommand{\GeV}{\  {\rm GeV} }
\newcommand{\TeV}{\  {\rm TeV} }

\newcommand{\lmk}{\left(}  
\newcommand{\rmk}{\right)}
\newcommand{\lkk}{\left[}  
\newcommand{\rkk}{\right]}

\newcommand{\bea}{\begin{array}}
\newcommand{\eea}{\end{array}}
\newcommand{\beq}{\begin{eqnarray}}
\newcommand{\eeq}{\end{eqnarray}}

\newcommand{\abs}[1]{\left\vert {#1} \right\vert}

\def\REF#1{Ref.~\cite{#1}}
\def\REFS#1{Refs.~\cite{#1}}
\def\SEC#1{Sec.~\ref{#1}}
\def\FIG#1{Fig.~\ref{#1}}
\def\EQ#1{Eq.~(\ref{#1})}

\newcommand{\h}{_{h}}
\newcommand{\bl}{_{(B - L)_{3}}}

\title{
Self-interacting dark matter with a vector mediator: kinetic mixing with U(1)$_{\bf (B-L)_3}$ gauge boson
}

\author{
Ayuki Kamada,$^{1}$
}
\affiliation{
$^{1}$ Center for Theoretical Physics of the Universe, 
Institute for Basic Science (IBS), Daejeon 34126, Korea
}

\author{
Masaki Yamada,$^{2}$
}
\affiliation{
$^{2}$ Institute of Cosmology, Department of Physics and Astronomy, 
Tufts University, Medford, MA 02155, USA
}

\author{
Tsutomu T. Yanagida$^{3, 4, 5}$
}
\affiliation{$^{3}$ Kavli IPMU (WPI), UTIAS, 
The University of Tokyo, 
Kashiwa, Chiba 277-8583, Japan
}
\affiliation{
$^{4}$ T. D.  Lee Institute and School of Physics and Astronomy, Shanghai Jiao Tong University, Shanghai 200240, China
}
\affiliation{
$^{5}$ Hamamatsu Professor
}

\abstract{
A spontaneously broken hidden U(1)$\h$ gauge symmetry can explain both the dark matter stability and the observed relic abundance. In this framework, the light gauge boson can mediate the strong dark matter self-interaction, which addresses astrophysical observations that are hard to explain in collisionless cold dark matter. Motivated by flavoured grand unified theories, we introduce right-handed neutrinos and a flavoured $B - L$ gauge symmetry for the third family U(1)$\bl$. The unwanted relic of the U(1)$\h$ gauge boson decays into neutrinos via the kinetic mixing with the U(1)$_{(B - L)_3}$ gauge boson. Indirect detection bounds on dark matter are systematically weakened, since dark matter annihilation results in neutrinos. However, the kinetic mixing between U(1)$_{(B - L)_3}$ and U(1)$_Y$ gauge bosons are induced by quantum corrections and leads to an observable signal in direct and indirect detection experiments of dark matter. This model can also explain the baryon asymmetry of the Universe via the thermal leptogenesis. In addition, we discuss the possibility of explaining the lepton flavour universality violation in semi-leptonic $B$ meson decays that is recently found in the LHCb experiment.
}

\begin{document}

\maketitle
\flushbottom

\section{Introduction
\label{sec:introduction}}

The nature of dark matter (DM) is a longstanding mystery in cosmology and particle physics. If DM consists of some new particle, it needs to be long-lived and its relic density should explain the observed amount. A simple framework to explain these aspects is to introduce a gauge symmetry U(1)$\h$, which is broken to some discrete group. The DM particle is stable due to the unbroken discrete group and the correct relic density is obtained through the DM annihilation into light gauge bosons $Z\h$. 

The light gauge boson mediates a DM self-interaction, which can address small-scale issues in structure formation of collisionless cold dark matter (see, e.g., \REFS{Tulin:2017ara, Bullock:2017xww}). The self-interaction is velocity-dependent as astrophysical observations prefer~\cite{Kaplinghat:2015aga}. The cross section per mass is $\sigma / m \gtrsim 1 \, {\rm cm}^2/{\rm g}$ in (dwarf) galaxies to explain, e.g., diversity in galaxy rotation curves~\cite{Kamada:2016euw, Creasey:2016jaq, Ren:2018jpt} (see also \REF{Oman:2015xda}). Meanwhile it diminishes to $\sigma / m \lesssim 0.1 \, {\rm cm}^2/{\rm g}$ in galaxy clusters to be compatible, e.g., with the inferred core of relaxed galaxy clusters~\cite{Kaplinghat:2015aga} (see also \REF{Newman:2012nw}). The circular velocity in galaxy clusters is of order $v \sim 1000 \, {\rm km / s}$, while that in dwarf galaxies is of order $v \sim 30 {\rm km / s}$. It implies that a velocity-dependent self-interaction is preferred. 

$Z\h$ tends to be stable, while one can make the U(1)$\h$-breaking scalar decay into two $Z\h$'s. Thermally produced $Z\h$ may overclose the Universe if it is stable. One may introduce a kinetic mixing between $Z\h$ and the U(1)$_Y$ hypercharge gauge boson so that $Z\h$ can decay into an electron-positron pair or photons. However, late-time DM annihilation followed by the $Z\h$ decay is largely disfavored by indirect detection constraints, e.g. from cosmic microwave background (CMB) anisotropies (see, e.g., \REF{Bringmann:2016din}). One way to avoid the overclosure of $Z\h$ and these constraints is to make it decay only into standard model (SM) neutrinos.

A similar line of constructing a viable self-interacting DM model was pursued in \REF{Kamada:2018zxi}, where U(1)$\h$ is identified as a flavoured lepton gauge symmetry U(1)$_{L_{\mu} - L_{\tau}}$. The MeV-scale $L_{\mu} - L_{\tau}$ gauge boson decays predominantly into neutrinos since charged lepton channels are kinematically forbidden. On the other hand, the gauge coupling needs to be rather small to satisfy constraints from muon anomalous magnetic moment and thus the mediated self-interaction is also small. This is why the MeV-scale U(1)$_{L_{\mu} - L_{\tau}}$-breaking scalar was considered as a scalar mediator of the DM self-interaction. In this paper, we propose a self-interacting DM model with a vector mediator~\cite{Feng:2009mn, Tulin:2012wi, Dasgupta:2013zpn, Bringmann:2013vra, Ko:2014bka, Cherry:2014xra, Kitahara:2016zyb, 
Ma:2017ucp, Balducci:2018ryj}. 

We consider flavoured U(1)$_{B - L}$ gauge symmetries; we introduce a $B - L$ gauge symmetry U(1)$_{(B - L)_{i}}$ for each family ($i = 1, 2, 3$) in the SM sector. The anomaly cancellation implies that there is a right-handed neutrino $N_{R_i}$ in each family. This model could be extended to grand unified theories such as $[SO(10)]^{3}$~\cite{Alonso:2017uky} (see also \REF{Babu:2007mb}). We assume that U(1)$\bl$ is spontaneously broken around the electroweak scale. MeV-scale $Z\h$ decays predominantly into neutrinos through kinetic mixing between Z$\h$ and the U(1)$\bl$ gauge boson $Z_{(B-L)_3}$ since channels into quarks and charged leptons are not kinematically allowed. In our model, we make the mass of the U(1)$\h$-breaking scalar larger than $2 \times m_{Z\h}$ so that the scalar field can decay into two $Z\h$'s. 
Quantum corrections give a kinetic mixing between U(1)$\bl$ and U(1)$_Y$ gauge bosons because there are bicharged particles in the SM sector. Because of these kinetic mixings, 
our DM is within reach of direct detection experiments in the near future. 
In addition, the kinetic mixings 
make the DM annihilation lead to an electron-positron pair with a small branching ratio.
Indirect detection experiments can potentially examine the signals.

Interestingly, we can realize the seesaw mechanism~\cite{Minkowski:1977sc, Yanagida:1979as, GellMann:1980vs, Glashow:1979nm} and the thermal leptogenesis~\cite{Fukugita:1986hr} (see, e.g., \REFS{Buchmuller:2002rq, Giudice:2003jh, Buchmuller:2005eh, Davidson:2008bu} for recent reviews) by assuming U(1)$_{(B - L)_{1}}$ and U(1)$_{(B - L)_{2}}$ to be spontaneously broken at the scale above $10^{9} \GeV$. We assume that electroweak-scale $N_{R_3}$ is stable because of a ${\mathbb Z}_{2}$ symmetry so as not to washout the $B - L$ asymmetry via its decay~\cite{Cox:2017rgn}. Our model can also explain the recent measurement of the lepton flavour universality violation in semi-leptonic $B$ meson decays~\cite{Aaij:2014ora, Aaij:2017vbb} because the U(1)$\bl$ gauge boson mediates flavour universality violating interactions for mass eigenstates of quarks and leptons.

This paper is organized as follows. First, we specify our model of DM and flavoured U(1)$_{(B-L)_i}$. We introduce a spontaneously broken U(1)$\h$ gauge symmetry in the dark sector. The U(1)$\bl$ is spontaneously broken around the electroweak scale so that the kinetic mixing with $Z\bl$ leads to the decay of $Z\h$ into SM neutrinos. In \SEC{sec:cosmology}, we discuss the cosmology of this model. In particular, we discuss that there are two candidates of DM in this model: the vector-like fermion in the hidden sector and $N_{R_3}$. The former one has the self-interaction through the massive gauge boson exchange and is assumed to be the dominant component of DM. Then, we discuss the compatibility with the present collider experiments and future detectability in \SEC{sec:collider}. \SEC{sec:conclusions} is devoted to the conclusion.

\section{Model
\label{sec:model}}

We introduce three right-handed neutrinos $N_{R_i}$ and flavoured U(1)$_{(B-L)_i}$ gauge symmetries ($i=1,2,3$) 
that are spontaneously broken by vacuum expectation values (VEVs) of $\Phi_i$ at the energy scale of $v_{\phi_i}$. 
We make the third family of right-handed neutrino stable by introducing a ${\mathbb Z}_{2}$ symmetry. 
We also introduce another complex scalar field $\Psi$ and a vector-like fermion pair $\chi$ and $\bar{\chi}$ that are charged under a hidden gauge symmetry U(1)$\h$.
The field $\Psi$ is assumed to obtain a nonzero VEV to break U(1)$\h$ spontaneously at the energy scale of $v_\psi$. 
The charge of $\Psi$ is taken to be three in units of that of $\chi$ to forbid Yukawa interactions with $\chi$ or $\bar{\chi}$. 
The charge assignment of the newly introduced particles is summarized in Table~\ref{table1}, 
where we omit the first and second families for simplicity.

The Lagrangian is given by 
\beq
&& {\cal L} = {\cal L}_{\rm SM} + 
 {\cal L}_{\rm kin} + 
 {\cal L}_{1,2} + 
 {\cal L}_{3} + 
 {\cal L}_{h} \,, \\
&& {\cal L}_{1,2} =  
 - \frac{1}{2} \sum_{i=1}^2 y_{R_i} \Phi_i N_{R_i} N_{R_i} 
 - \sum_{i=1}^2 y_{R}^{i} H N_{R_i} L_i + {\rm h.c.}
 - \sum_{i=1}^2 V_{\Phi_i} (\Phi_i) \,, \\
&& {\cal L}_3 =  
 - \frac{1}{2} y_{R_3} \Phi_3 N_{R_3} N_{R_3} + {\rm h.c.} 
 - V_{\Phi_3} (\Phi_3) \,, 
\\
&& {\cal L}\h =  
 - m_\chi \chi \bar{\chi}  + {\rm h.c.}
 - V_\Psi (\Psi) \,,
\eeq
where ${\cal L}_{\rm SM}$ and ${\cal L}_{\rm kin}$ represent the SM Lagrangian and canonical kinetic terms of the newly introduced particles including the gauge interactions, respectively. $H$ and $L_{i}$ are the SM Higgs and lepton doublets, respectively. $y_{R_{i}}$ and $y_{R}^{i}$ ($y_{R}^{3} = 0$) are dimensionless Yukawa couplings.

The scalar fields are assumed to be unstable at the origins of the potentials, $V_{\Phi_{i}} (\Phi_{i})$ and $V_{\Psi} (\Psi)$, and 
obtain nonzero VEVs $v_{\phi_i}$ and $v_\psi$ at the stable minima. 
We denote the perturbations around the minima as $\phi_i$ and $\psi$ as 
\beq
 &&\Phi_i = \frac{1}{\sqrt{2}} ( v_{\phi_i} + \phi_i) \,,
 \\
 &&\Psi = \frac{1}{\sqrt{2}} ( v_\psi + \psi) \,. 
\eeq 
After the spontaneous symmetry breaking (SSB), 
the gauge bosons $Z_{(B-L)_i}$ and $Z\h$ obtain masses such as 
$m_{Z_{(B-L)_i}} = 2 g_{(B-L)_i} v_{\phi_i}$ and 
$m_{Z_{h}} = 3 g_{h} v_{\psi}$ with $g_{(B - L)_{i}}$ and $g_{h}$ being the gauge couplings, respectively. 
The right-handed neutrinos obtain masses of $m_{N_{R_i}} = y_{R_i} v_{\phi_i} / \sqrt{2}$ 
via the Yukawa interaction. 
The SM neutrinos obtain small masses via the seesaw mechanism. 

{\renewcommand\arraystretch{1.2}
\begin{table}[t]
\begin{center}
\caption{Charge assignment.
\label{table1}}
\begin{tabular}{p{2cm}p{1.5cm}p{1.5cm}p{1.5cm}p{1.5cm}p{0.75cm}}
\hline
\hline
& $N_{R_3}$ & $\Phi_3$ & $\chi$ &$ \bar{\chi} $& $\Psi$ \\
\hline
U(1)$\bl$ & $-1$ & $2$  & $0$ & $0$ & $0$ \\
U(1)$\h$ & $0$ & $0$ & $1$ & $-1$ & $3$  \\
${\mathbb Z}_{2}$ & $-1$ & $+1$  & $+1$ & $+1$ & $+1$ \\
\hline \hline
\end{tabular}\end{center}
\end{table}
}

Because of the flavoured symmetry, 
the proper structure of Yukawa interactions cannot be generated in the simplest setup. 
To generate the proper Yukawa matrices, one may introduce (I) U(1)$\bl$-charged scalars in addition to U(1)$\bl$-neutral vector-like fermions that mix with the SM quarks and leptons~\cite{Alonso:2017uky}; or (II) an additional Higgs doublet that is charged under U(1)$\bl$~\cite{Bian:2017rpg, Dev:2017xry, Duan:2018akc}.
The additional fields lead to additional collider constraints on the model; e.g., vector-like fermions should be heavier than TeV~\cite{Alonso:2017uky}.
We do not go into further detail about ultraviolet setups.
We assume that after integrating out the heavy ($\sim \TeV$) fields, low-energy phenomenology is described by the above Lagrangian with the following Yukawa structure. 

The SM Yukawa matrices can be diagonalized by a unitary rotation for each fermion: $f = U_{f} f'$ ($f = u_{L}, d_{L}, u_{R}, d_{R}, \nu_{L}, l_{L}, l_{R}$). 
Although each unitary matrix is not observable in the SM, except for $U^\dagger_{u_L} U_{d_L} = V_{\rm CKM}$ and $U_{l_L}^\dagger U_{\nu_L} = U_{\rm PMNS}$, it affects the interactions with the $Z\bl$ boson.
The interactions with the $Z\bl$ boson are given by 
\beq
 &&{\cal L} \supset - \sum_f g_{(B-L)_3} Q_f Z\bl^\mu J_{f, \mu} \,,
 \\
 &&J_{f, \mu} = \sum_{i,j = 1}^3 \bar{f}_i (U_f)_{3i}^* (U_f)_{3 j} \gamma_\mu f_j \,. 
\eeq
$Z\bl$ can mediate interactions between different families 
in the mass eigenstate 
even if only the third family fermions are charged under U(1)$_{(B-L)_3}$ in the interaction basis. 

In this paper, we simply assume that the rotations of the right-handed fermions are suppressed and the $2$-$3$ family rotations of the left-handed fermions exist in addition to $V_{\rm CKM}$ and $U_{\rm PMNS}$ 
such as 
\beq
\label{eq:UeL}
 &&U_{l_L} = R_{23} (\theta_l) \,, ~~~~~~~~~~~~~ 
 U_{\nu_L} = R_{23} (\theta_l) U_{\rm PMNS} \,, 
 \\
 &&U_{d_L} = R_{23} (\theta_q) \,, ~~~~~~~~~~~~~ 
 U_{u_L} = R_{23} (\theta_q) V_{\rm CKM}^\dagger \,,
\eeq
where $R_{23}(\theta)$ is a 2-3 family rotation by an angle $\theta$.
In particular we assume that $R_{13}(\theta)$ does not arise so that $Z\h$ does not decay into electrons via the kinetic mixing with $Z\bl$.

\section{Cosmology of the model
\label{sec:cosmology}}

\subsection{Thermal leptogenesis}

We can generate the lepton asymmetry by the thermal leptogenesis via the decay of 
the first and second family right-handed neutrinos. 
We assume that the reheating temperature after the inflation is higher than the mass of the lighter one among these right-handed neutrinos 
so that they can be produced from the thermal plasma. 
The lepton asymmetry can be generated by their decay. 
Since the $B+L$ symmetry is broken by the non-perturbative effect, 
we can generate the baryon asymmetry from the lepton asymmetry. 
The observed baryon asymmetry can be explained when the lighter one is heavier than about $10^{9} \GeV$~\cite{Fukugita:1986hr}. 

If the third family right-handed neutrino has a Yukawa interaction with the SM particles, 
the $B-L$ symmetry violating interaction may be in equilibrium after the thermal leptogenesis 
and the lepton asymmetry may be washed out. 
To avoid this washout effect, 
we impose a ${\mathbb Z}_{2}$ symmetry on $N_{R_3}$. 
As a result, it is stable and can be a DM candidate. 

If we do not introduce the ${\mathbb Z}_{2}$ symmetry on $N_{R_3}$, the Yukawa coupling with the SM fields $y_{R}^{3}$ should be small enough to suppress the washout effect. 
The decay rate of $N_{R_3}$ is given by 
\beq
 \Gamma_{N_{R_3}} \simeq \frac{\abs{y_{R}^{3}}^2}{8 \pi} m_{N_{R_3}} \,.
\eeq
The washout effect should not be efficient, $\Gamma_{N_{R_3}} \lesssim H$, 
until the temperature of the Universe decreases to the mass of $N_{R_3}$. 
Thus we require 
\beq
 \sqrt{\sum_i \abs{y_{R}^{3i}}^2} \lesssim 2 \times 10^{-7} \lmk \frac{m_{N_{R_3}}}{1 \TeV} \rmk^{1/2} \,, 
\eeq
to avoid the washout effect.

\subsection{Dark matter}

There are two DM candidates in our model. 
We identify $\chi$ and $\bar{\chi}$ as the dominant component of DM, 
while $N_{R_3}$ is the subdominant component.
Their thermal relic densities are determined as 
\beq
 \Omega_i h^2 \approx 0.12 \lmk \frac{3 \times 10^{-26} {\rm \ cm}^3 {\rm s}^{-1}}{( \sigma_i v )} \rmk \,, 
 \label{DMabundance}
\eeq
with the $s$-wave annihilation cross section times relative velocity $( \sigma_i v )$.

\subsubsection{Weakly-interacting DM: $N_{R_3}$}

The annihilation of $N_{R_3}$ proceeds through the U(1)$\bl$ gauge interaction 
and the Yukawa interaction with $\Phi_3$. 
We found in \REF{Cox:2017rgn} that 
the dominant process is a $s$-wave annihilation channel $N_{R_3} N_{R_3} \to Z\bl \phi_3$ 
if it is kinematically allowed and $m_{N_{R_3}} \gg m_{Z\bl}, m_{\phi_3}$. 
The cross section is given by 
\beq
 &&(\sigma_{N_{R_3}} v) (N_{R_3} N_{R_3} \to Z\bl \phi_3) 
 \nonumber\\
 &&\simeq \frac{\pi \alpha\bl^2 }{4 m_{N_{R_3}}^4 m_{Z\bl}^4} 
\lkk m_{\phi_3}^4 - 2 m_{\phi_3}^2 ( 4 m_{N_{R_3}}^2 + m_{Z\bl}^2) + (4 m_{N_{R_3}}^2 - m_{Z\bl}^2)^2 \rkk^{3/2} \,, 
 \nonumber\\
 \label{annihilation}
\eeq
where $\alpha\bl \equiv g\bl^2 / (4\pi)$. 
The resulting amount of $N_{R_3}$ can be then estimated as 
\beq
 \Omega_{N_{R_3}} h^2 \approx 1.4 \times 10^{-2} \lmk \frac{m_{N_{R_3}}}{1 \ {\rm TeV}} \rmk^{-2} \lmk \frac{m_{Z\bl}}{70 \ {\rm GeV}} \rmk^{4} \lmk \frac{\alpha\bl}{10^{-4}} \rmk^{-2} \,, 
 \label{Omega_L}
\eeq
for $m_{N_{R_3}} \gtrsim m_{Z\bl}, m_{\phi_3}$. 
We assume that $N_{R_3}$ is the subdominant component of DM: $\Omega_{N_{R_3}} h^2 \ll (\Omega_{\rm DM} h^2)^{\rm obs} \approx 0.12$. 
Then we obtain 
\beq
 y_{R_3}^{-1} v_{\phi_3} \ll 2 \TeV \,. 
 \label{constraintYukawa}
\eeq

The Yukawa coupling $y_{R_3}$ cannot be arbitrary large because of the Unitarity bound. 
One may also require that the Landau pole does not appear below the Planck scale, 
which leads to $y_{R_3} \lesssim 1.2$~\cite{Cox:2017rgn}. 
Then we obtain $v_{\phi_3} \ll 2.4 \TeV$ from \EQ{constraintYukawa}.

Although the dominant annihilation channel is $s$-wave and its cross section is not suppressed at in the late Universe, $N_{R_3}$ is not the dominant component of the DM and hence its indirect detection signals can be neglected.

\subsubsection{Self-interacting DM: $\chi$ and $\bar{\chi}$}
For $\chi$ and $\bar{\chi}$, 
the annihilation cross section is given by~\cite{Duerr:2018mbd}
\beq
 &&(\sigma_\chi v) (\chi \bar{\chi} \to Z\h Z\h) \simeq \frac{\pi \alpha\h^2}{m_\chi^2} \,, 
 \\
 &&(\sigma_\chi v) (\chi \bar{\chi} \to Z\h \psi) \simeq \frac{9\pi \alpha\h^2}{4m_\chi^2} \,. 
\eeq
where $\alpha\h \equiv g\h^2 / (4 \pi)$.
The total abundance of $\chi$ and $\bar{\chi}$ is twice larger than \EQ{DMabundance} because there are $\chi$ and $\bar{\chi}$, each abundance of which is determined by the thermal freeze out.
The resulting amount of $\chi$ and $\bar{\chi}$ can be then estimated as%
\footnote{Depending on $m_{Z\h}$, the Sommerfeld enhancement can be significant~\cite{Binder:2017lkj}.
When we focus on the regime where a large self-scattering cross section alleviates small-scale issues, $m_{\chi} \lesssim 100$\,GeV is free from this subtlety.}
\beq
 \Omega_\chi h^2 \approx 0.13 \lmk \frac{m_{\chi}}{40 \ {\rm GeV}} \rmk^{2} \lmk \frac{\alpha\h}{10^{-3}} \rmk^{-2} \,. 
 \label{Omega_chi}
\eeq

The massive gauge boson $Z\h$ mediates the self-interaction of $\chi$ and $\bar{\chi}$. 
It is convenient to use the transfer cross section defined by~\cite{Tulin:2013teo}%
\footnote{
\label{footnote2}
Replacing $1- \cos \theta$ by $1- |\cos \theta|$ is suggested since backward scattering has nothing to do with phase space redistribution as forward scattering~\cite{Kahlhoefer:2013dca, Kahlhoefer:2017umn}.}
\beq
 &&\sigma_T \equiv \frac12 \lmk \sigma_T^{(\chi \chi)} + \sigma_T^{(\chi \bar{\chi})} \rmk \,, 
 \\
 &&\sigma_T^{(\chi \chi), (\chi \bar{\chi})} = \int d \Omega ( 1- \cos \theta) \lmk \frac{d \sigma^{(\chi \chi), (\chi \bar{\chi})}}{d \Omega} \rmk \,. 
 \label{sigmaT}
\eeq 
When one computes $\sigma_T$, one encounters three regimes~\cite{Tulin:2013teo}: Born regime ($\alpha\h m_\chi / m_{Z\h} \ll 1$), classical regime ($\alpha\h m_\chi / m_{Z\h} \gtrsim 1$ \& $m_\chi v_{\rm rel} / m_{Z\h} \gg 1$), and resonance regime ($\alpha\h m_\chi / m_{Z\h} \gtrsim 1$ \& $m_\chi v_{\rm rel} / m_{Z\h} \lesssim 1$).
In the Born regime, one can rely on the perturbative calculation and find an analytic expression in \REFS{Feng:2009hw, Tulin:2013teo, Kahlhoefer:2017umn}.
In the classical and resonance regimes, one needs to solve the Schr\"odinger equation to take into account non-perturbative effects related to multiple exchanges of $Z\h$.
Meanwhile, fitting formulas can be found in the classical regime~\cite{PhysRevLett.90.225002, PhysRevLett.90.225002, Feng:2009hw, Tulin:2013teo, Cyr-Racine:2015ihg}.
In the resonance regime, an approximate formula can be obtained in the Hulth\'en potential~\cite{Tulin:2013teo}.

Kinematics of dwarf and low-surface brightness galaxies indicate that 
$\sigma / m_{\chi} \approx 1$-$10 \, {\rm cm^{2}/g}$ for the DM velocity of order $30 \, {\rm km/s}$~\cite{Kaplinghat:2015aga}. 
On the other hand, observations of galaxy clusters prefer $\sigma_T/ m_\chi \lesssim 0.1 \, {\rm cm}^2/{\rm g}$ for the velocity of order $1000 \, {\rm km/s}$~\cite{Kaplinghat:2015aga}.
If the cross section saturates this upper bound, we can also explain the inferred density cores in the galaxy clusters~\cite{Newman:2012nw}.
The desirable parameter region is mostly in the resonance regime (see, e.g., Ref.~\cite{Kamada:2018zxi}), where the parameter dependence of the self-scattering cross section is non-trivial. 
In this paper, 
we do not pin down the precise values of $m_\chi$ and $m_{Z\h}$ 
because they are not sensitive to other observables. 
Instead,
we simply use an approximate formulas found in \REF{Tulin:2013teo} 
with the replacement of $\cos \theta \to \abs{\cos \theta}$ in \EQ{sigmaT} (see footnote~\ref{footnote2}) 
to check if the self-interaction cross section is within a desirable range. 
We find that 
the above constraints can be satisfied when 
$m_{Z\h} \approx 10 \text{-} 100 \MeV$ 
and $m_\chi \approx 10 \text{-} 100 \GeV$.

\subsection{Dark radiation}

We assume that the mass of the $U(1)_{h}$ gauge boson $\psi$ is larger than twice that of $Z\h$, 
so that it can decay into two $Z\h$'s. 
We make $Z\h$ unstable by introducing a kinetic mixing between U(1)$\bl$ and U(1)$\h$:
\beq
 - \frac12 \epsilon_2 F\bl^{\mu \nu} F_{\h \, \mu \nu} \,, 
\eeq 
where $F\bl$ and $F\h$ denote the field strengths of U(1)$\bl$ and U(1)$\h$, respectively. 
Then $Z\h$ can decay into third family neutrinos $\nu_{3}$ via the mixing with $Z\bl$.
We remark that the decay of $Z\h$ into muons $\mu$ or taus $\tau$ is kinematically forbidden for $m_{Z\h} \lesssim 200 \MeV$.
The other decay of $Z\h$ into electrons $e$ is suppressed since $Z\bl$ does not directly couple to $e$ under our assumption of the Yukawa structure [see \EQ{eq:UeL}].
Thus the late time DM annihilation into $Z\h$ results in $\nu_3$ and thus is harmless.

The decay rate can be estimated as 
\beq
 \Gamma_{Z\h} \sim \alpha\bl \epsilon_2^2 \lmk \frac{m_{Z\h}}{m_{Z\bl}} \rmk^4 m_{Z\h} \,.
\eeq
We require that $Z\h$ decays into $\nu_3$ long before the neutrino decoupling; otherwise only the temperature of $\nu_3$ is enhanced by the decay of $Z\h$
and the energy density of $\nu_3$ may exceed the upper bound on that of dark radiation. 
This can be satisfied when $\Gamma_{Z\h} \gtrsim H \vert_{T =1\MeV}$, where $H$ is the Hubble expansion rate at temperature $T$.
It gives the lower bound on the mixing parameter as 
\beq
 \epsilon_2 \gtrsim 
 4 \times 10^{-2} \lmk \frac{m_{Z\bl}}{70 \GeV} \rmk^{2} 
 \lmk \frac{m_{Z\h}}{10 \MeV} \rmk^{-5/2} 
 \lmk \frac{\alpha\bl}{10^{-4}} \rmk^{-1/2} \,. 
\label{epsilon2}
\eeq
Even if $Z\h$ decays into $\nu_3$ long before the neutrino decoupling, the thermalized $Z\h$ can still enhance only the temperature of $\nu_3$ after the neutrino decoupling.
This constraint is evaded for $m_{Z\h} \gtrsim 10 \MeV$~\cite{Kamada:2015era, Kamada:2018zxi}.

Furthermore, one needs to take account of $Z\h$ possibly dominating the energy density of the Universe.
It takes place if the decay rate of the $U(1)_{h}$ gauge boson is much smaller than the Hubble expansion rate when the temperature is comparable to the mass of dark Higgs boson, $\Gamma_{Z\h} \lesssim H \vert_{T = m_{Z\h}}$.
If $m_{Z\h} > 1 \MeV$ and $\Gamma_{Z\h} \lesssim H \vert_{T =1 \MeV}$, the Hubble expansion rate during the big bang nucleosynthesis is dominated by non-relativistic $Z\h$ and affects the big bang nucleosynthesis critically.%
\footnote{ This point seems missing in Ref.~\cite{Duerr:2018mbd}, where the lightest particle in a hidden sector ($Z\h$ in our case) is the dark Higgs boson.
They take about $1.5 \MeV$ as a reference value of the dark Higgs boson mass and require its lifetime to be shorter than $10^{5} \ {\rm s}$ not to affect the CMB spectral distortion (see, e.g., \REF{Poulin:2016anj}). 
However, the lifetime of the dark Higgs boson should be shorter than ${\cal O}(1) \ {\rm s}$ not to dominate the energy density of the Universe 
if the dark sector is decoupled from the SM sector after the QCD phase transition. 
In Ref.~\cite{Hufnagel:2018bjp}, they have investigated this effect in detail 
and found that the lifetime can be as long as ${\cal O}(100) \ {\rm s}$ 
for the ${\cal O} (1) \, {\rm MeV}$ dark Higgs boson if it is decoupled before the QCD phase transition. 
}
If $m_{Z\h} > 1 \MeV$ and $\Gamma_{Z\h} \gtrsim H \vert_{T =1\MeV}$, the $Z\h$ domination does not impact the big bang nucleosynthesis, but still dilutes the baryon asymmetry.
To avoid such a wash out, the lower bound of the mixing should satisfy
\beq
 \epsilon_2 \gtrsim 
 4 \times 10^{-1} \lmk \frac{m_{Z\bl}}{70 \GeV} \rmk^{2} 
 \lmk \frac{m_{Z\h}}{10 \MeV} \rmk^{-3/2}
 \lmk \frac{\alpha\bl}{10^{-4}} \rmk^{-1/2} \,. 
\label{epsilon3}
\eeq
If this condition is not satisfied, 
the amount of the entropy production due to the $Z\h$ decay can be estimated as 
\beq
\Delta &\equiv& \frac{s_{f} a_{f}^{3}}{s_{i} a_{i}^{3}} 
\simeq 
 \frac{m_{Z\h}}{T_d} 
 \\
 &\approx& 10 \lmk \frac{\epsilon_2}{4 \times 10^{-2}} \rmk^{-1} 
 \lmk \frac{m_{Z\bl}}{70 \GeV} \rmk^{2} 
 \lmk \frac{m_{Z\h}}{10 \MeV} \rmk^{-3/2}
 \lmk \frac{\alpha\bl}{10^{-4}} \rmk^{-1/2} \,, 
\eeq
where 
$a_{i} (a_{f})$ and $s_{i} (s_{f})$ are the scale factor and entropy density before (after) the $Z_{h}$ domination, respectively, and 
$T_d$ is the decay temperature of $Z\h$. 
The constraint (\ref{epsilon3}) can be evaded if the generated baryon asymmetry is 
larger than the observed value by this factor. 
This can be realized when the first and second right-handed neutrinos are heavier than $10^9 \GeV$ at least by the same factor.

Here we comment on another possible mechanism of the entropy production, which could be relevant in models with a spontaneous symmetry breaking. 
As for a dynamics of U(1)$\h$ breaking in the hidden sector, 
a thermal inflation may occur at the time of the phase transition 
if the mass of the gauge boson $m_{Z\h}$ is many orders of magnitude larger than 
that of the symmetry-breaking field $m_\psi$. 
This effect washes out the baryon asymmetry, so that we should avoid such a thermal inflation. 
We discuss the condition to avoid a thermal inflation in Appendix~\ref{sec:appendix} 
and check that it does not occur in our model. 
However, we note that it is non-trivial in other models with hierarchical mass scales.

\subsection{DM direct and indirect detection constraints}

There may be couplings between scalar fields like 
$\lambda_{H \Phi_i} \abs{H}^2 \abs{\Phi_i}^2$ 
and $\lambda_{H \Psi} \abs{H}^2 \abs{\Psi}^2$. 
Since they are irrelevant in the above discussion, we take the loop induced values as natural choices.
For example, the former interaction arises at the two loop level as $\lambda_{H \Phi_3} \sim y_t^2 \alpha\bl^2 / (4 \pi)^2$.
It results in the mixing between the SM Higgs and $\phi$, 
which leads to spin-independent $N_{R_3}$-nucleon scatterings. 
However, $N_{R_3}$ is the subdominant component of DM 
and hence easily evades the constraint from the direct detection experiments for DM. 
For the same reason, the indirect detection constraint on $N_{R_3}$ is also weakened. 

The kinetic mixing between the U(1)$_Y$ and U(1)$\bl$ gauge bosons arises at the one-loop level: 
\beq
 &&{\cal L}_{\rm kin} \supset - \frac12 \epsilon_1 F_{Y \, \mu \nu} F\bl^{\mu \nu} \,,
 \\
 &&\epsilon_1 \simeq \frac{2 g_Y g\bl}{9 \pi^2} \ln \lmk \frac{\Lambda}{\mu} \rmk
 \approx 10^{-2} \lmk \frac{\alpha\bl}{10^{-4}} \rmk^{1/2} 
  \ln \lmk \frac{\Lambda}{10^{16} \GeV} \frac{10^{2} \GeV}{\mu} \rmk  
 \,,
 \label{epsilon1}
\eeq
where $F_{Y \, \mu \nu}$ and $g_{Y}$ are the field strength and gauge coupling for U$(1)_{Y}$, respectively.
Here $\mu$ is the energy scale considered 
and $\Lambda$ is a cutoff scale at which the kinetic mixing vanishes. 
When U(1)$_Y$ is unified into a non-abelian gauge symmetry, 
we should take $\Lambda$ to be the grand unification scale of order $10^{16} \GeV$.%
\footnote{
We implicitly consider SU(5)$\times$U(1)$\bl \times$U(1)$\h$ gauge theory as an effective theory, where SU(5) breaks down to the SM gauge groups at the GUT scale. 
In this case, the kinetic mixing between 
the U(1)$_Y$ and U(1)$\h$ gauge bosons are forbidden above the GUT scale 
while the one between the U(1)$\bl$ and U(1)$\h$ gauge bosons is allowed by the symmetry, SU(5)$\times$U(1)$\bl \times$U(1)$\h$. 
Although the former one can be induced after the GUT symmetry breaking, it depends on the detail of the model. In this paper, we assume that 
the mixing parameter is suppressed enough so that our DM can evade the constraints coming from DM direct detection experiments. 
}
Through the kinetic mixings parametrized by $\epsilon_1$ and $\epsilon_2$, 
we obtain effective interactions between $\chi$ and SM particles as 
calculated in Appendix~\ref{sec:mixing}. 
In particular, 
there is the following effective interaction at a low energy scale: 
\beq
 {\cal L} \supset b_p \bar{\chi} \gamma^\mu \chi \bar{p} \gamma_\mu p \,, 
\eeq
where $p$ represents the proton. 
From Eq.~(\ref{interactionM}) and the discussion below the equation, 
we estimate the coefficient $b_{p}$ roughly as
\beq
 b_p \sim 
 \frac{e g\h \cos \theta_W \epsilon_1 \epsilon_2 }{m_{Z_{(B-L)_3}}^2 } 
 f(q^2/m_{Z_h}^2)\,, 
\eeq
where $e$ is the electromagnetic charge,
$q^2$ is the squared momentum transfer, 
and 
\beq
 f(x) = \frac{x}{1+ x} \,. 
\eeq
The momentum transfer is of order $\mu v_{\rm DM}$, 
where $\mu$ ($\sim m_\chi$) 
is the reduced mass for $\chi$ and the nucleus 
and $v_{\rm DM}$ ($\sim 10^{-3}$) is the relative velocity. 
Noting that $m_{Z_h} / m_\chi \sim 10^{-3}$, 
we expect $f(q^2/m_{Z_h}^2) = {\cal O}(1)$.

For a given nucleus $^A_Z N$, 
the coefficient of the coupling is given by $b_N = Z b_p$, 
where we neglect the contribution comes from the neutron. 
Then the spin-independent $\chi$-nucleon scattering cross section is given by 
\beq
 \sigma_N 
 &=& \frac{1}{\pi} \frac{\mu_N^2}{A^2} b_N^2 
 \\
 &\sim& 
 7 \times 10^{-48} \, {\rm cm}^2 \times f(q^2/m_{Z_h}^2) 
 \lmk \frac{\alpha\h}{10^{-3}} \rmk 
 \lmk \frac{\epsilon_1}{10^{-2}} \rmk^2
 \lmk \frac{\epsilon_2}{10^{-2}} \rmk^2
 \lmk \frac{m_{Z\bl}}{70 \GeV} \rmk^{-4} \,,
\eeq
(see, e.g., Ref.~\cite{Fitzpatrick:2010em}), where $\mu_N$ is the $\chi$-nucleon reduced mass. 
This is just below the present upper bound reported by XENON1T 
for $m_\chi = 20$-$100 \GeV$~\cite{Aprile:2018dbl}. 
XENONnT~\cite{Aprile:2015uzo}, DarkSide-20k~\cite{Aalseth:2017fik}, and LUX-ZEPLIN~\cite{Mount:2017qzi} can search DM with a cross section smaller by a factor of order $10$. 
DARWIN can detect DM if the cross section is above 
the neutrino coherent scattering cross section~\cite{Aalbers:2016jon}, 
which is around $10^{-49} \, {\rm cm}^2$ for $m_\chi = 20$-$100 \GeV$.

As we stressed, the late-time annihilation of the dominant component of DM, $\chi$ and $\bar{\chi}$, 
predominantly results in $\nu_3$. 
Since the detection of neutrino signals is quite challenging and the constraint is very weak~\cite{Frankiewicz:2015zma}, this does not lead to observable effects on astrophysical experiments. 
However, 
their annihilation can also result in an electron-positron pair via the process of $Z_h \to e \bar{e}$. 
The branching ratio is calculated from Eq.~(\ref{interactionM}) and the result is given by 
\beq
 {\rm Br} \lmk Z_h \to e \bar{e} \rmk 
 &\simeq& 2 c_W^2 \epsilon_1^2 \frac{\alpha}{\alpha_h} 
 \\
 &\approx& 10^{-3}  \lmk \frac{\epsilon_1}{10^{-2} } \rmk^2 \lmk \frac{\alpha\h}{10^{-3}} \rmk^{-1} \,,
\eeq
where $\alpha = e^{2} / (4 \pi)$.
The effective annihilation cross section of DM into electron-positron pairs is given by 
\beq
 (\sigma_\chi v) (\chi \bar{\chi} \to e \bar{e} \cdots) 
 &\simeq& 
S \lkk 2 {\rm Br} \times (\sigma_\chi v) (\chi \bar{\chi} \to Z_h Z_h) 
 + {\rm Br} \times  (\sigma_\chi v) (\chi \bar{\chi} \to Z_h \psi) \rkk 
 \\
 &\approx&
 1.1 \times 10^{-28} \, {\rm cm}^3/ s 
 \, S \lmk \frac{\epsilon_1}{10^{-2} } \rmk^2 \lmk \frac{m_{\chi}}{40 \ {\rm GeV}} \rmk^{2} \lmk \frac{\alpha\h}{10^{-3}} \rmk^{-3} \,, 
\eeq
where $S$ represents a Sommerfeld enhancement factor. 
An upper bound on the annihilation cross section into an electron-positron pair is obtained from AMS-02 data as ${\cal O} (1) \times 10^{-26} \, {\rm cm^{3}/s}$ for $m_{\chi} = {\cal O} (10) \, {\rm GeV}$ ($\approx 1 \times 10^{-26} \, {\rm cm^{3}/s}$ for $m_{\chi} =40 \, {\rm GeV}$)~\cite{Leane:2018kjk}.%
\footnote{In Ref.~\cite{Leane:2018kjk}, they consider the case of Majorana DM while we consider Dirac DM. To compare the result, we multiply the upper bound by a factor of 2.}
We calculate the Sommerfeld enhancement factor by using the Hulth\'en potential as done in Ref.~\cite{Feng:2010zp}. 
The result is typically ${\cal O} (1)$ for $v \sim 100 \, {\rm km / s}$ ($S \approx 3$ for $m_{\chi} = 40 \, {\rm GeV}$, $\alpha_{h} = 10^{-3}$, and $m_{Z_{h}} = 10 \, {\rm MeV}$) .
The Sommerfeld enhancement factor increases toward lower velocity and saturate at 
\beq
S \approx 50 \lmk \frac{m_{\chi}}{40 \, {\rm GeV}} \rmk \lmk \frac{\alpha_{h}}{10^{-3}} \rmk \lmk \frac{m_{Z_{h}}}{10 \, {\rm MeV}} \rmk^{-1}
\eeq
unless a parameter is tuned to enhance it resonantly.
The resultant cross section is close to an upper bound from CMB anisotropies~[55], ${\cal O} (1) \times 10^{-25} \, {\rm cm^{3}/s}$ ($\approx 8 \times 10^{-26} \, {\rm cm^{3}/s}$ for $m_{\chi} =40 \, {\rm GeV}$).
In summary, the present constraints on indirect detection are already constraining some parameter space.
We expect that indirect detection experiments with a large exposure and a better understanding of cosmic ray background and propagation will examine a broader parameter space.

\section{Collider constraints
\label{sec:collider}}

As discussed in \SEC{sec:model}, we assume that the appropriate flavour structure of the SM Yukawa matrices comes from some UV physics (see, e.g., Ref~\cite{Alonso:2017uky}).
The CKM and PMNS matrices are attributed to the left-handed up-type quarks and neutrinos, respectively.
We allow for additional $2$-$3$ family rotations of left-handed quarks and leptons.

\subsection{No additional physical phase
\label{sec:flavour0}}
First we discuss the constraints from collider experiments, when there are no additional physical phases, $\theta_{l} = \theta_{q} = 0$, following \REF{Cox:2017rgn}. 

A relevant constraint on the $Z\bl$ mass comes from the lepton flavour universality violation in $\Upsilon$ decays. 
The lepton flavour universality ratio is modified in the presence of $Z\bl$ as 
\beq
 R_{\tau \mu} ( \Upsilon (1S))
 &\equiv& \frac{\Gamma_{ \Upsilon (1S) \to \tau {\bar \tau}}}{\Gamma_{ \Upsilon (1S) \to \mu {\bar \mu}}}
 \\
 &\simeq& \lmk 1+ \frac{\alpha\bl}{\alpha} \frac{m_\Upsilon^2}{m_{Z\bl}^2 - m_\Upsilon^2}\rmk^2 \,, 
\eeq
where $m_\Upsilon$ ($\approx 9.46 \GeV$) is the Upsilon mass. 
The BaBar experiment places the constraint on this ratio as $R_{\tau \mu} = 1.005 \pm 0.013 ({\rm stat.}) \pm 0.022({\rm syst.})$~\cite{delAmoSanchez:2010bt}. 
In the limit of $m_{Z\bl}^2 \gg m_\Upsilon^2$, 
we obtain 
\beq
 \lmk \frac{m_{Z\bl}}{70\GeV} \rmk^2 \lmk \frac{\alpha\bl}{10^{-4}} \rmk^{-1} \gtrsim 1.7 \times 10^{-2} \,, 
\eeq
corresponding to
\beq
 v_{\phi_3} \gtrsim 130 \GeV \,. 
 \label{upsilon}
\eeq
This is consistent with the upper bound on $v_{\phi_3}$ from the $N_{R_3}$ abundance [see Eq.~(\ref{constraintYukawa})].

Since the U(1)$_{(B-L)_3}$ gauge boson 
is coupled with the third-family quarks, 
it can be produced in hadron colliders. 
The dominant production process is a Drell-Yan process 
from the bottom quark pair: $b \bar{b} \to Z\bl$. 
Resonance searchs in the $\tau \bar{\tau}$ final state place the constraint for $200 \GeV \lesssim m_{Z\bl} \lesssim 4 \TeV$~\cite{Aaboud:2017sjh}.
This constraint, $\alpha\bl \lesssim 10^{-3}$ for $m_{Z\bl} \sim 200 \GeV$, would not be quite stringent when compared to others.%
\footnote{Our reference value of $m_{Z\bl}$ is $70 \GeV$, which is slightly out of this range.
We expect that the constraint does not change drastically (see also Refs.~\cite{Sirunyan:2017nvi, Aaboud:2018zba}).}

\begin{figure}[t] 
   \centering
   \includegraphics[width=3.5in]{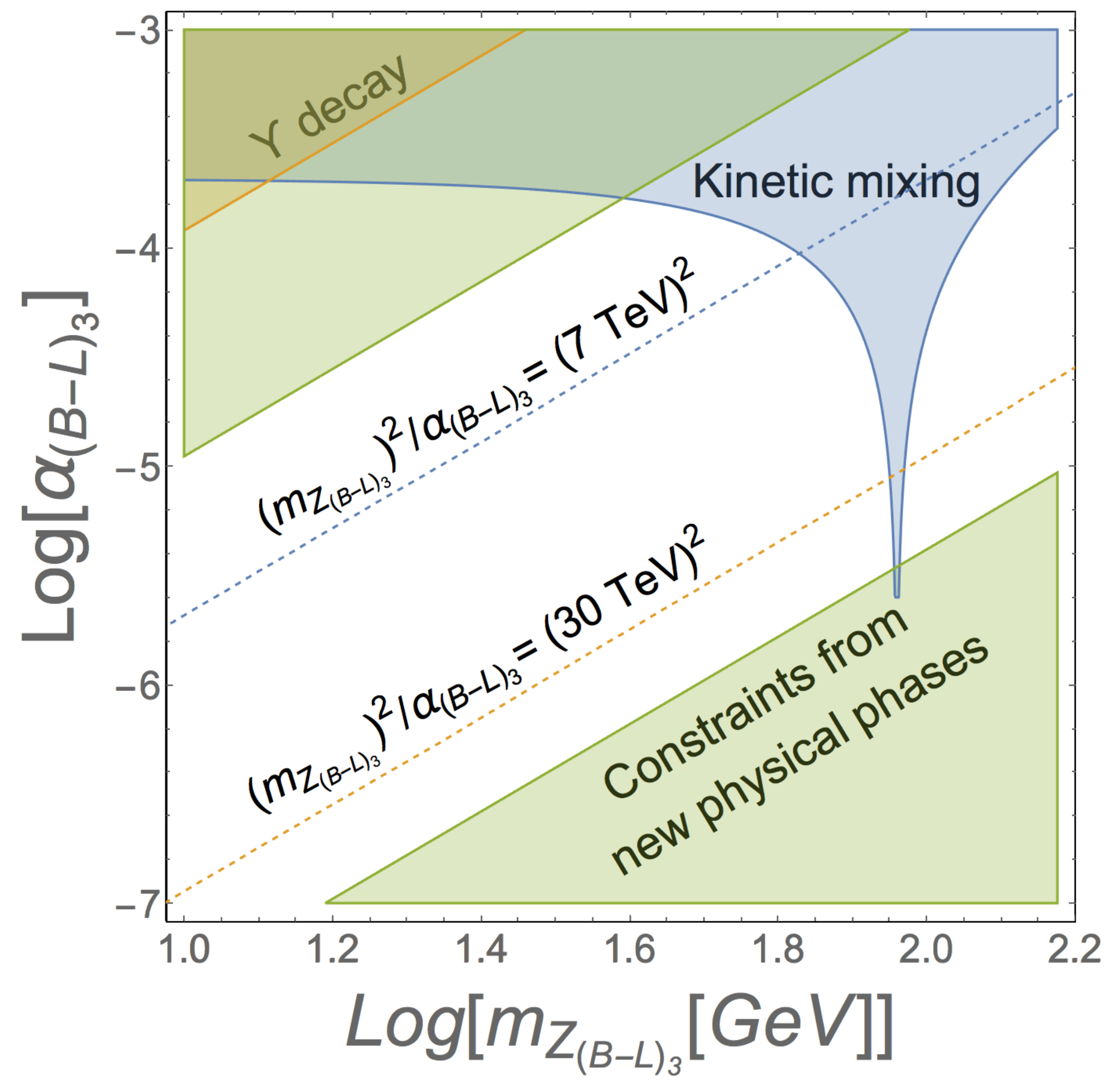} 
   \caption{
Parameter region in the $\alpha\bl$-$m_{Z\bl}$ plane.
The orange and blue shaded regions are constrained by the $\Upsilon$ decay and 
kinetic mixing, respectively,
when $\theta_{l} = \theta_{q} = 0$.
In the green shaded regions, one cannot find a region in the $\sin \theta_{l}$-$\sin \theta_{q}$ plane, where the flavour constraints are satisfied.
The dashed lines are the reference parameter values that we use in \FIG{fig2}. 
   }
   \label{fig1}
\end{figure}

The kinetic mixing between the $Z\bl$ and U(1)$_Y$ gauge bosons changes the mass eigenstates and interactions of the vector mesons.
In particular it leads to a shift in the SM $Z$ boson mass. 
In the mass basis, the physical mass of the SM $Z$ boson, denoted by $m_{Z_1}$, is given by Eq.~(\ref{m_Z1}). 
The mass of the SM $Z$ boson is tightly constrained 
by electroweak precision measurements 
and is consistent with the SM prediction. 
In this paper we require that the mass difference is smaller than the current experimental uncertainty of $0.0021 \GeV$~\cite{Tanabashi:2018oca}. 
The kinetic mixing parameter $\epsilon_1$ can be written in terms of $\alpha\bl$ 
as in \EQ{epsilon1}. 
Then we 
can plot a constraint in the $\alpha\bl$-$m_{Z\bl}$ plane as shown by the blue region in \FIG{fig1}. 
The orange region in the upper left corner is excluded by the $\Upsilon$ decay measurement. 
We also plot a green region that is excluded by the flavour physics as we will discuss in \SEC{sec:flavour}. 
The dashed lines are a couple of reference parameter values that we will use in \FIG{fig2}.

The Higgs-portal interaction $\lambda_{H \Phi_3} \abs{H}^2 \abs{\Phi_3}^2$ may provide indirect signals of Higgs invisible decays 
if the SM Higgs can decay into $N_{R_3} N_{R_3}$, $Z\bl Z\bl$, or $\phi_3 \phi_3$. 
The constraint, however, can be easily evaded unless the Higgs-portal coupling is as large as ${\cal O}(1)$.

\subsection{Non-zero new physical phases}
\label{sec:flavour}

Next we allow the additional physical phases to be non-zero.
The following discussion is based on \REF{Alonso:2017uky}.

\subsubsection{Semi-leptonic $B$ decays}
To study semi-leptonic $B$ decays, it is convenient to use the following effective Hamiltonian 
at the low energy: 
\beq
 &&{\cal H}_{\rm eff} = - \frac{4 G_F}{\sqrt{2}} V_{tb} V_{ts}^* 
 \lkk \sum_{l = e, \mu, \tau} \lmk C_9^l {\cal O}_9^l + C_{10}^l {\cal O}_{10}^l \rmk +\sum_{i, j = 1}^{3}  C_\nu^{ij} {\cal O}_\nu^{ij} \rmk \,,
 \\
 &&{\cal O}_9^l = \frac{\alpha}{4 \pi} \lmk \bar{s} \gamma_\mu b_L \rmk \lmk 
 \bar{l} \gamma_\mu l \rmk \,, 
 \\
 &&{\cal O}_{10}^l = \frac{\alpha}{4 \pi} \lmk \bar{s} \gamma_\mu b_L \rmk \lmk \bar{l} \gamma_\mu \gamma^5 l \rmk \,, 
 \\
 &&{\cal O}_\nu^{ij} = \frac{\alpha}{2 \pi} \lmk \bar{s} \gamma_\mu b_L \rmk \lmk \bar{\nu}_i \gamma_\mu \nu_{L \, j} \rmk \,. 
\eeq

After integrating out $Z\bl$, 
we obtain the deviation from the SM contributions for $\mu$ such as 
\beq
 \delta C_9^\mu = - \delta C_{10}^\mu = - \frac{\pi}{\alpha \sqrt{2} G_F V_{tb} V_{ts}^*} 
 \frac{g\bl^2 s_{\theta_q} c_{\theta_q} s_{\theta_l}^2}{3 m_{Z\bl}^2 } \,, 
\eeq
where $V_{tb}$ denotes the $tb$ component of $V_{\rm CKM}$ and so on.
Hereafter, we also use $s_{\theta_{q}} \equiv \sin \theta_{q}$ and so on.
The LHCb experiment reported the lepton flavour universality violation in semi-leptonic $B$ decays~\cite{Aaij:2014ora, Aaij:2017vbb}, 
which is represented by the ratio of 
\beq
 {\cal R}_K^{(*)} = \frac{\Gamma (B \to K^{(*)} \mu {\bar \mu}) }{\Gamma (B \to K^{(*)} e {\bar e}) } \,. 
\eeq
The tension with the SM prediction is around the $4\sigma$ level~\cite{Altmannshofer:2017yso, DAmico:2017mtc, Capdevila:2017bsm, Hiller:2017bzc, Ciuchini:2017mik, Geng:2017svp, Alok:2017sui}. 
This can be explained by the $Z\bl$ contribution 
when $\delta C_9^\mu \in [-0.81, - 0.48]$ ($1 \sigma$ interval)~\cite{CMS:2014xfa}. 
Using $\abs{V_{tb}} \approx 1.0$ and $\abs{V_{ts}} \approx 3.9 \times 10^{-2}$~\cite{Tanabashi:2018oca}, 
we then require 
\beq
 8.7 \times 10^{-3} \lesssim s_{\theta_q} c_{\theta_q} s_{\theta_l}^2 \lmk \frac{\alpha\bl}{10^{-4}} \rmk \lmk \frac{m_{Z\bl}}{70 \GeV} \rmk^{-2} \lesssim 1.5 \times 10^{-2} \,. 
 \label{Bdecay1}
\eeq

The $B$ meson can decay also into neutrinos: 
$B \to K^{(*)} \nu \bar{\nu}$. 
The deviation from the SM contribution is given by 
\beq
&& \delta C_\nu^{ij} = \delta C_\nu \lmk U_{\nu_L} \rmk_{3i}^* \lmk U_{\nu_L} \rmk_{3j} \,,
 \\
&& \delta C_\nu = - \frac{\pi}{\alpha \sqrt{2} G_F V_{tb} V_{ts}^*} 
 \frac{g\bl^2 s_{\theta_q} c_{\theta_q}}{3 m_{Z\bl}^2 } \,. 
\eeq
The ratio to the SM prediction is given by 
\beq
 R_{\nu \bar{\nu}} \equiv \frac{\Gamma}{\Gamma_{\rm SM}} = 1 
 + \frac{2}{3} \lmk \frac{\delta C_\nu}{C_\nu^{(\rm SM)}} \rmk 
 + \frac13 \lmk \frac{\delta C_\nu}{C_\nu^{(\rm SM)}} \rmk^2 \,, 
\eeq
where $C_\nu^{(\rm SM)} \approx - 6.35$~\cite{Buras:2014fpa}. 
The experimental upper bound is $R_{\nu \bar{\nu}} < 4.3$ at the 90\% confidence level (CL)~\cite{Lutz:2013ftz, Lees:2013kla}, 
which gives 
\beq
 s_{\theta_q} c_{\theta_q} \lmk \frac{\alpha\bl}{10^{-4}} \rmk \lmk \frac{m_{Z\bl}}{70 \GeV} \rmk^{-2} \lesssim 2.6 \times 10^{-1} \,. 
 \label{Bdecay2}
\eeq
Combining this with \EQ{Bdecay1}, 
we obtain 
\beq
 \abs{s_{\theta_l}} \gtrsim 0.18 \,. 
 \label{sthetal1}
\eeq

\subsubsection{$B_s$-$\bar{B}_s$ and $D^0$-$\bar{D}^0$ mixings}

$Z\bl$ also contributes to the $B_s$-$\bar{B}_s$ mixing 
via the following effective Lagrangian: 
\beq
 {\cal L} \supset - \frac{g\bl^2 s_{\theta_q}^2 c_{\theta_q}^2}{18 m_{Z\bl}^2} 
 \lmk \bar{s} \gamma^\mu b_L \rmk^2 \,. 
\eeq
This gives a deviation of the $B$ meson mass difference from the SM prediction as 
\beq
 C_{B_s} \equiv \frac{\Delta m_{B_s}}{\Delta m_{B_s}^{(\rm SM)}} 
 = 1 + \frac{4 \pi^2 c (m_{Z\bl})}{G_F^2 m_W^2 V_{tb} V_{ts}^* \hat{\eta}_B S(m_t^2/m_W^2)} 
 \frac{g\bl^2 s_{\theta_q}^2 c_{\theta_q}^2}{18 m_{Z\bl}^2} \,, 
\eeq
where $c(m_{Z\bl}) \approx 0.8$~\cite{Ciuchini:1997bw, Buras:2000if}, 
$S(m_t^2/m_W^2) \approx 2.30$~\cite{doi:10.1143}, and $\hat{\eta}_B \approx 0.84$~\cite{BURAS1990491, Lenz:2010gu}. 
The experimental constraint is $0.899 < C_{B_s} < 1.252$ ($95\%$ CL interval)~\cite{Bona:2007vi}. 
Then we obtain 
\beq
 \abs{s_{\theta_q} c_{\theta_q}} \lmk \frac{\alpha\bl}{10^{-4}} \rmk^{1/2} \lmk \frac{m_{Z\bl}}{70 \GeV} \rmk^{-1} \lesssim 2.0 \times 10^{-1} \,. 
 \label{mixing1}
\eeq
This upper bound is comparable to that in \EQ{Bdecay2}, in the parameter region that we are interested in.

The $D^0$-$\bar{D}^0$ mixing is induced by the $Z\bl$ exchange effective interaction of 
\beq
 {\cal L} \supset - \frac{g\bl^2 c_D^2}{18 m_{Z\bl}^2} 
 \lmk \bar{u} \gamma^\mu c_L \rmk^2 \,, 
\eeq
where 
\beq
 c_D \equiv \lmk V_{ub} c_{\theta_q} - V_{us} s_{\theta_q} \rmk 
 \lmk V_{cb}^* c_{\theta_q} - V_{cs}^* s_{\theta_q} \rmk \,. 
\eeq
This results in a new physics contribution of 
\beq
 \Delta m_D^{\rm NP} = \frac23 f_D^2 B_D m_D \, c(m_{Z\bl}) \, \frac{g\bl^2 c_D^2}{18 m_{Z\bl}^2} \,, 
\eeq
where $f_D \approx 207.4 \MeV$~\cite{Carrasco:2014poa} and $B_D \approx 0.757$~\cite{Carrasco:2015pra} are calculated by the lattice quantum chromodynamics. 
The $D^0$ meson mass is $m_D \approx 1.86 \GeV$. 
The mass difference calculated in the SM has large uncertainties~\cite{Golowich:2007ka}, 
so that we cannot evaluate the total (SM\,+\,NP) mass difference robustly.
In this paper we simply require that the new physics contribution 
does not exceed the experimental data, which is $4 \times 10^{-4} < \Delta m_D^{\rm NP}/\Gamma < 6.2 \times 10^{-3}$, 
where $\Gamma \approx 2.44 \times 10^{12} / {\rm s}$ is the average decay width of $D^0$ and $\bar{D}^0$~\cite{Amhis:2016xyh, Tanabashi:2018oca}. 
Since $\abs{V_{ub}} \approx 4.1 \times 10^{-3}$, 
$\abs{V_{us}} \approx 0.22$, 
$\abs{V_{cb}} \approx 4.1 \times 10^{-2}$, and 
$\abs{V_{cs}} \approx 1.0$, this constraint is stringent particularly for $c_{\theta_q} \ll 1$.
Then we obtain 
\beq
 \abs{s_{\theta_q}} \lmk \frac{\alpha\bl}{10^{-4}} \rmk^{1/4} \lmk \frac{m_{Z\bl}}{70 \GeV} \rmk^{-1/2} \lesssim 1.5 \times 10^{-1} \,
 \label{mixing1}
\eeq
for $s_{\theta_q} \gg 0.04 c_{\theta_q}$.

\subsubsection{Lepton flavour violation}

Lepton flavour violating processes are also induced by $Z\bl$ interactions. 
The most important effective interaction is 
\beq
 {\cal L} \supset \frac{g\bl^2}{m_{Z\bl}^2} s_{\theta_l}^3 c_{\theta_l} \bar{\tau} \gamma^\rho \mu_L \bar{\mu} \gamma_\rho \mu_L \,, 
\eeq
which gives the $\tau$ decay into $3 \mu$. 
The resulting branching ratio is given by 
\beq
 {\cal B}r (\tau \to 3 \mu) = \frac{m_\tau^5}{48 \pi \Gamma_\tau} 
 \frac{\alpha\bl^2}{m_{Z\bl}^4} s_{\theta_l}^6 c_{\theta_l}^2 \,, 
\eeq
where $m_\tau$ ($\approx 1.78 \GeV$) and $\Gamma_\tau \approx (2.9 \times 10^{-13} \, {\rm s})^{-1} \approx 2.3 \times 10^{-12} \GeV$ are the mass and decay width of the tau lepton, respectively. 
The experimental upper bound is ${\cal B}r (\tau \to 3 \mu) < 2.1 \times 10^{-8}$ at the $90\%$ CL~\cite{Hayasaka:2010np}. 
Thus we obtain 
\beq
 \abs{s_{\theta_l}^3 c_{\theta_l}} \lmk \frac{\alpha\bl}{10^{-4}} \rmk \lmk \frac{m_{Z\bl}}{70 \GeV} \rmk^{-2} 
 \lesssim 3.1 \times 10^{-2} \,. 
 \label{sthetal2}
\eeq
For the reference parameter values, $\alpha\bl = 10^{-4}$ and $m_{Z\bl} = 70 \GeV$, 
this constraint implies that 
$\abs{s_{\theta_l}} \lesssim 0.32$, which is compatible with \EQ{sthetal1}.

\subsubsection{$Z\bl$ production and decay in colliders}

The mixing in the lepton sector leads to new decay channels of $Z\bl$: $Z\bl \to \mu \bar{\mu}$ and $Z\bl \to \mu \tau$. 
We can place a constraint by using the dimuon search by the ATLAS and CMS collaborations at 
the LHC experiment for $200 \GeV < m_{Z\bl} < 4 \TeV$~\cite{Aaboud:2017buh, Aaboud:2018jff}.
The constraint is similar to $Z\bl \to \tau \bar{\tau}$ discussed in \SEC{sec:flavour0} and would not be quite stringent.

The SM $Z$ boson can decay into $\mu \bar{\mu} Z\bl$ followed by $Z\bl \to \mu \bar{\mu}$. 
The ATLAS collaboration reported an upper bound on the branching ratio of $Z \to 4 \mu$ as ${\cal B}r ( Z \to 4 \mu) < [3.2 \pm 0.25 ({\rm stat.}) \pm 0.13 ({\rm syst.})] \times 10^{-6}$~\cite{Aad:2014wra}. 
This can be interpreted as an upper bound on $g\bl s_{\theta_l}^4$ for a given $m_{Z\bl}$, where $g\bl s_{\theta_l}^2$ comes from a coupling for the $Z\bl$ production process and another $s_{\theta_l}^2$ comes from a branching ratio of $Z\bl$ decay into $\mu \bar{\mu}$.%
\footnote{
The $Z\bl$ boson can also decay into $\mu \bar{\mu}$ via the kinetic mixing with U(1)$_Y$. Since this process does not dominate for $s_{\theta_l}$ satisfying (\ref{sthetal1}), we neglect this contribution. 
}
We simply estimate the constraint by replacing $g'$ of Fig.~1 of \REF{Bonilla:2017lsq} with our $g\bl s_{\theta_l}^4/3^2$, where a factor of $3^2$ comes from a difference of the charge between the models. 
The resulting constraint is $g\bl s_{\theta_l}^4 \lesssim 6 \times 10^{-2}$ for $m_{Z\bl} = 20$-$30 \GeV$, 
while it is two orders of magnitude weaker for $m_{Z\bl} \gtrsim 70 \GeV$. 
We find that this constraint is negligible in most of the parameter region that we are interested in.

Because of the kinetic mixing, 
$Z\bl$ can be produced by hadron colliders via the Drell-Yan process, 
which leads to a clear dilepton signal 
with an invariant mass about the $Z\bl$ boson mass. 
In Ref.~\cite{Hoenig:2014dsa}, 
they considered the case 
where a hidden gauge boson is coupled with the SM sector only via the kinetic mixing 
and estimated that 
the $8\TeV$ LHC with $20 \, {\rm fb}^{-1}$ luminosity puts
a constraint on the kinetic mixing parameter 
as $\epsilon_1 \lesssim 0.005$-$ 0.01$ for the gauge boson mass range of $10$-$70 \GeV$. 
In our model, 
the $Z\bl$ boson can be produced via the kinetic mixing%
\footnote{
The $Z\bl$ boson can also be produced from $b \bar{b}$, 
in which case the constraint can be interpreted as a bound on 
$f g\bl / e s_{\theta_{l}}^2$, where $f = {\cal O} (\alpha_{s})$, with the quantum chromodynamics fine-structure constant $\alpha_{s}$, represents a factor coming from parton distribution functions of $b$ and $\bar{b}$. 
}
and decay into $\mu \bar{\mu}$ via the flavour mixing. 
The constraint can be interpreted as a bound on $\epsilon_1 s_{\theta_l}^2$, where $s_{\theta_l}^2$ comes from the branching ratio into $\mu \bar{\mu}$. However, this does not give a strong constraint on $s_{\theta_l}$ 
when $\alpha\bl \lesssim 10^{-4}$. 
It was also discussed that 
the constraint will be improved by 
a factor about $5$ by using $3000 \, {\rm fb}^{-1}$ of $14 \TeV$ data. 
In this case, the high-luminosity LHC would observe a dilepton signal 
for $\alpha\bl \sim 10^{-4}$.

The $Z\bl$ boson can also be produced by lepton colliders through the kinetic mixing $\epsilon_{1}$ 
and its decay signal can be searched by the future $e {\bar e}$ colliders, 
such as CEPC~\cite{CEPCStudyGroup:2018rmc}, ILC~\cite{Baer:2013cma, Fujii:2017vwa}, 
and FCC-ee~\cite{Gomez-Ceballos:2013zzn}. 
The relevant process is $e {\bar e} \to \gamma Z\bl$ followed by $Z\bl \to \mu {\bar \mu}$. 
Projected constraints are discussed in Ref.~\cite{He:2017zzr} 
in the case where the dark photon couples to the SM particles only via the kinetic mixing with the U(1)$_Y$ gauge boson. 
The upper bound on the kinetic mixing parameter was found to be about $0.003$. 
Again, we could interpret their result in the same way as discussed above. 
The future lepton colliders would observe a signal of $Z\bl$ boson 
for $\alpha\bl \sim 10^{-4}$.

\subsection{Summary of the collider constraints}

\begin{figure}[t] 
   \centering
   \includegraphics[width=2.5in]{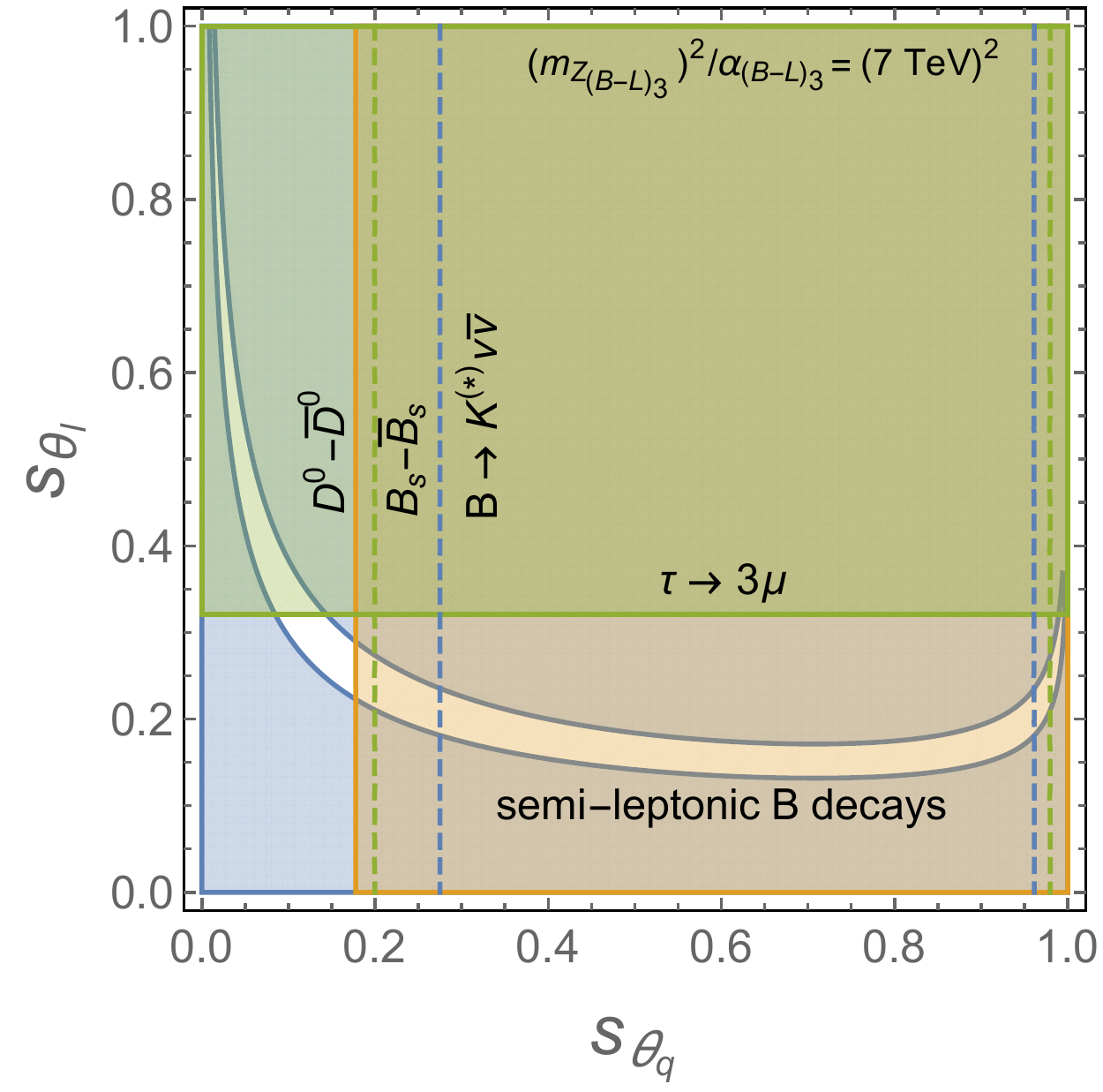} 
   \quad
   \includegraphics[width=2.5in]{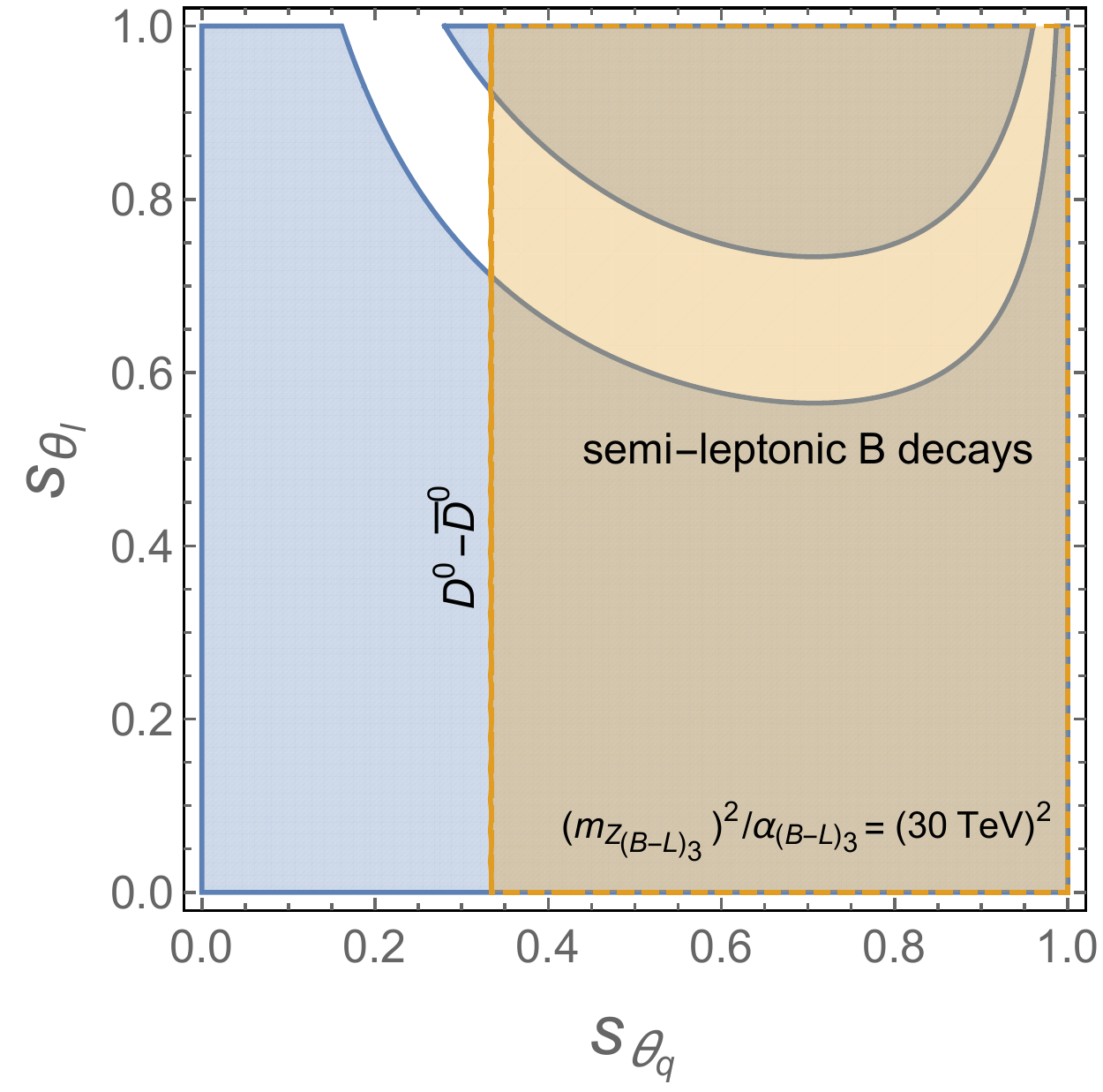} 
   \caption{
Constraints are summarized in the $s_{\theta_l}$-$s_{\theta_q}$ plane, 
where the shaded region is excluded by the experiments. 
All the displayed constraints depend only on the combination of $m_{Z\bl}^2 / \alpha\bl$, 
which is taken to be $(7 \TeV)^2$ (left panel) and $(30 \TeV)^2$ (right panel). 
   }
   \label{fig2}
\end{figure}

Now we shall put together all the constraints discussed in this section. 
The result is shown in \FIG{fig2}, 
where the shaded regions are excluded by the constraints. 
Note that 
all the shown constrains depend on $\alpha\bl$ and $m_{Z\bl}$ 
only via a combination of $m_{Z\bl}^2 / \alpha\bl$. 
In the figure, we take $m_{Z\bl}^2 / \alpha\bl = (7 \TeV)^2$ (left panel) 
or $(30 \TeV)^2$ (right panel).

We change the value of $m_{Z\bl}^2 / \alpha\bl$ 
and find that there is an allowed region 
when $(3 \TeV)^2 \lesssim m_{Z\bl}^2 / \alpha\bl \lesssim (49 \TeV)^2$ 
corresponding to $0.4 \TeV \lesssim v_{\phi_3} \lesssim 6.9 \TeV$. 
This is shown in \FIG{fig1} as the green-shaded region. 
From \FIG{fig1}, we can see that there is a certain parameter region where we can explain the lepton flavour universality violation in semi-leptonic $B$ decays consistently with the constraints coming from the kinetic mixing. 
We note that $v_{\phi_{3}} \ll 2.4 \, {\rm TeV}$ is required so that $N_{R_{3}}$ is the subdominant component of DM [see Eq.~(\ref{constraintYukawa})].
In particular, all collider constraints as well as the DM constraints 
can be evaded when 
$\alpha_{(B-L)_3} = 10^{-4}$, 
$\alpha_h = 10^{-3}$, 
$m_{Z_{(B-L)_3}} = 70 \GeV$, 
$m_{Z_h} = 10 \MeV$, 
$m_\chi = 40 \GeV$, 
$\epsilon_1 = 10^{-2}$ 
and $\epsilon_2 = 4 \times 10^{-2}$, 
which are used as the reference parameter values throughout this paper.

\section{Conclusion
\label{sec:conclusions}}

We have proposed a model of DM whose stability is guaranteed by a discrete symmetry. This discrete symmetry is a subgroup of a spontaneously broken hidden U(1)$_h$ gauge symmetry. The massive gauge boson $Z\h$ is assumed to be much lighter than the DM particle and mediates the velocity-dependent DM self-interaction that are suggested by small-scale issues in structure formation of collisionless cold dark matter. The observed abundance of DM is explained by the thermal relic via the freeze-out mechanism. Motivated by flavoured grand unified theories, we have also introduced right-handed neutrinos and flavoured $B-L$ gauge symmetries. The unwanted relic of $Z\h$ can then decay into neutrinos via the kinetic mixing with the electroweak scale U(1)$_{(B - L)_3}$ gauge boson $Z\bl$. This model can also explain the baryon asymmetry of the Universe via the thermal leptogenesis. Furthermore, we have found that this model can explain the lepton flavour universality violation in semi-leptonic $B$ meson decays recently found in LHCb experiment. 

Although the hidden sector couples to the SM sector only via the kinetic mixing with the U(1)$_{(B - L)_3}$ gauge boson, it predicts detectable DM signals in 
direct detection experiments. 
Furthermore the subdominant DM annihilation into an electron-positron pair can be examined in indirect detection experiments. 
Our model predicts a relatively light U$(1)_{(B - L)_{3}}$ gauge boson, which leads to interesting signals in collider phenomenology. The U$(1)_{(B - L)_{3}}$ gauge boson can be searched by the future high-luminosity LHC experiment and $e {\bar e}$ colliders such as CEPC, ILC, and FCC-ee.
These experiments would observe signals 
when the fine-structure constant for U(1)$_{(B - L)_3}$ is of order $10^{-4}$.

%
\section*{Acknowledgments}
%

A.~K. thanks Tae Hyun Jung, Takumi Kuwahara, Chan Beom Park, and Eibun Senaha for useful discussions.
A.~K. was supported by Institute for Basic Science under the project code, IBS-R018-D1.
T.~T.~Y. was supported in part by the JSPS Grant-in-Aid for Scientific Research No. 26104001, No. 26104009, No. 16H02176, and No. 17H02878, and in part by World Premier International Research Center Initiative (WPI Initiative), MEXT, Japan. T.~T.~Y. is a Hamamatsu Professor at Kavli IPMU.

\appendix

\section{Mixing of gauge bosons}
\label{sec:mixing}

In this Appendix, we calculate the mixings among the gauge bosons and interactions induced by the kinetic mixing terms. 

The relevant part of the Lagrangian is given by 
\beq
 &&{\cal L}_{\rm gauge} = 
   -\frac{1}{4} F_{Y \mu\nu} F_Y^{\mu\nu}- \frac{1}{4} F_{W^{3} \, \mu \nu} F_{W^{3}}^{\mu \nu} 
   -\frac{1}{4} F_{(B-L)_3\, \mu\nu} F_{(B-L)_3}^{\mu\nu}
  -\frac{1}{4} F_{h\, \mu\nu} F_{h}^{\mu\nu} \nonumber \\
  &&~~~~~~~~~~ -\frac{1}{2} \sin\epsilon_1 F_{Y \, \mu\nu} F_{(B-L)_3}^{\mu\nu} 
  -\frac{1}{2} \cos \epsilon_1 \sin\epsilon_2 F_{(B-L)_3 \, \mu\nu} F_{h}^{\mu\nu} 
  -\frac{1}{2}  V^{\rm T}_\mu M^2_V V^\mu  \, ,
\eeq
where we have replaced $\epsilon_1$ and $\epsilon_2$ by $\sin \epsilon_1$ ($\simeq \epsilon_1$ for $\epsilon_1 \ll 1$)
and $\cos \epsilon_1 \sin \epsilon_2$ ($\simeq \epsilon_2$ for $\epsilon_1, \epsilon_2 \ll 1$) for convenience. 
Here $F_{W^{3} \, \mu \nu}$ is the field strength of the Cartan subgroup of weak SU$(2)$. We have defined $V^{\mu} = (B^{\mu}, W^{3 \, \mu}, Z^{\mu}_{(B - L)_{3}}, Z^{\mu}_{h})^{\rm T}$ with $B^{\mu}$ and $W^{3 \, \mu}$ being the gauge bosons of U$(1)_{Y}$ and the Cartan subgroup of weak SU$(2)$, respectively. The mass squared matrix is given by 
\begin{equation}
M^2_V=
\begin{pmatrix}
  m^2_Z s^2_W & -m^2_Z c_W s_W & 0 & 0 \\ 
  -m^2_Z c_W s_W & m^2_Z c^2_W & 0 & 0 \\  
  0 & 0 &    m^2_{Z_{(B-L)_3}} & 0 \\
  0 & 0  & 0 &    m^2_{Z_h} \\
\end{pmatrix}\, , 
\end{equation}
where $m_Z = m_{W^\pm} / c_W$, 
$c_W \equiv \cos \theta_W$, and $s_W \equiv \sin \theta_W$ 
with $m_{W^{\pm}}$ and $\theta_{W}$ being the $W$ boson mass and Weinberg angle, respectively. 

First, we diagonalize the kinetic terms and isolate the massless gauge boson $A^{\mu}$ by 
\begin{align}
  \begin{pmatrix}
    B^\mu \\ W^{3 \, \mu} \\ Z_{(B-L)_3}^\mu \\ Z_h^\mu
  \end{pmatrix}
  &= 
    \begin{pmatrix}
      1\ & 0\ & -t_{\epsilon_1}/c_{\epsilon_2}  & 0
      \\ 0\ & 1\ & 0 & 0
      \\  0\ & 0\ & 1/(c_{\epsilon_1} c_{\epsilon_2}) & 0 
      \\ 0\ & 0\ & -t_{\epsilon_2}  & 1
    \end{pmatrix} 
  \begin{pmatrix}
      c_W & -s_W & 0\ & 0
      \\ s_W & c_W & 0\  & 0 
      \\  0 & 0 & 1\ & 0 
      \\  0 & 0 & 0\ & 1       
    \end{pmatrix} 
    \begin{pmatrix}
      A^\mu \\ \tilde{Z}_{1}^{\mu} \\ \tilde{Z}_{2}^{\mu} \\ \tilde{Z}_{3}^{\mu}
    \end{pmatrix} \, ,
\end{align}
where $t_{\epsilon_{i}} \equiv \tan \epsilon_{i}$ and $c_{\epsilon_{i}} \equiv \cos \epsilon_{i}$ ($i = 1, 2, 3$). 
In terms of this basis, 
the mass squared matrix for $({\tilde Z}_{1}^{\mu}, {\tilde Z}_{2}^{\mu}, {\tilde Z}_{3}^{\mu})$ is given by ${\tilde M}_{V}^{2}$, where 
\beq
&&    (\tilde{M}_V^2 )_{11} = m^2_Z  \, , 
\\
&&    (\tilde{M}_V^2 )_{12} = m^2_Z s_W t_{\epsilon_1} /c_{\epsilon_2}\, ,
\\
&&    (\tilde{M}_V^2 )_{13} = 0\, , 
\\
&&    (\tilde{M}_V^2 )_{22} = m^2_{Z_{(B-L)_3}}/(c^2_{\epsilon_1}c^2_{\epsilon_2}) + m^2_Z s^2_W t^2_{\epsilon_1} /c_{\epsilon_2}^2 + m_{Z_h}^2 t_{\epsilon_2}^2\, ,
\\
&&    (\tilde{M}_V^2 )_{23} = - m^2_{Z_h} t_{\epsilon_2} \, ,
\\
&&    (\tilde{M}_V^2 )_{33} = m^2_{Z_h}\, . 
\eeq

Second, we diagonalize the mass squared matrix by rotating the vector fields as 
\begin{align}
  \begin{pmatrix}
      A^\mu \\ \tilde{Z}_{1}^{\mu} \\ \tilde{Z}_{2}^{\mu} \\ \tilde{Z}_{3}^{\mu}
  \end{pmatrix}
  &= 
  R_{12} (\xi_1) R_{32}(\xi_2) R_{31} (\xi_3) 
     \begin{pmatrix}
      A^\mu \\ Z_{1}^{\mu} \\ Z_{2}^{\mu} \\ Z_{3}^{\mu}
    \end{pmatrix} \, ,
\end{align}
where $R_{ij}(\xi)$ is the rotation matrix for $(Z_{i}^{\mu}, Z_{j}^{\mu})^{\rm T}$ ($i, j = 1, 2, 3$) by an angle $\xi$: 
\beq
&&R_{12} (\xi_1) = 
\begin{pmatrix}
      1 & 0 & 0 & 0 
      \\ 0 & c_{\xi_1} & -s_{\xi_1}  & 0 
      \\  0 & s_{\xi_1} &    c_{\xi_1} & 0 
      \\ 0 & 0 & 0 & 1 
    \end{pmatrix}\, , 
    \\
&&R_{32} (\xi_2) = 
  \begin{pmatrix}
      1\ & 0\ & 0 & 0 
      \\ 0\ & 1\ & 0 & 0 
      \\ 0\ & 0\ & c_{\xi_2} & s_{\xi_2}  
      \\  0\ & 0\ & -s_{\xi_2} &    c_{\xi_2} 
    \end{pmatrix}\, , 
\\
&&R_{31} (\xi_3) = 
\begin{pmatrix}
      1 & 0 & 0 & 0
      \\ 0 & c_{\xi_3} & 0  & s_{\xi_3}
      \\  0 & 0 &    1& 0 
      \\ 0& -s_{\xi_3}  & 0 & c_{\xi_3}
    \end{pmatrix}\, .    
\eeq
In this paper, we are interested in the case where $\epsilon_i \ll 1$ and $m_{Z_h} \ll m_{Z_{(B-L)_3}}, m_{Z}$. 
In this case, we can approximate the rotation angles as 
\beq
&&\tan2\xi_1 
\simeq \frac{2 (\tilde{M}_V^2 )_{12}}{(\tilde{M}_V^2 )_{11} - (\tilde{M}_V^2 )_{22} } 
\simeq \frac{2 s_{W} \epsilon_{1} m_{Z}^{2}}{m_{Z}^{2} - m_{(B - L)_{3}}^{2}}\, , 
\\
&& 
\xi_2 
\simeq \frac{ (\tilde{M}_V^2 )_{23} c_{\xi_1}}{ (\tilde{M}_V^2 )_{33} - m_{Z'_2}^2} 
\simeq c_{\xi_1} \epsilon_2 \frac{m_{Z_h}^2 }{ m_{Z'_2}^2}\, , 
\\
&& 
\xi_3 
\simeq \frac{ (\tilde{M}_V^2 )_{23} s_{\xi_1} c_{\xi_2} }{ (\tilde{M}_V^2 )_{33} - m_{Z_1}^2} 
\simeq s_{\xi_1} \epsilon_2 \frac{m_{Z_h}^2 }{ m_{Z_1}^2}\, , 
\eeq
and the mass eigenvalues as 
\beq
&&  m^2_{Z_{1}}
  \simeq
  \frac{m_Z^2 - m_{Z'_2}^2 s_{\xi_1}^2}{c_{\xi_1}^2}\, , 
  \label{m_Z1}
  \\
&&  m^2_{Z_{2}} \simeq m^2_{Z'_{2}} \simeq 
 c_{\xi_1}^2 m_{(B-L)_3}^2 
+ s_{\xi_1}^2 m_Z^2 \lmk 1 - \frac{2 s_W \epsilon_1}{t_{\xi_1} } \rmk\, , 
\\
&&  m^2_{Z_{3}} 
  \simeq 
  m_{Z_h}^2\, , 
\eeq
in the leading order for $\epsilon_i$ and $m_{Z_h} / m_{Z_{(B-L)_3}}$, 
where we have defined%
\footnote{
There are some typos in the literature in the context of kinetic mixing between U(1)$_Y$ and U(1)$'$. 
In Ref.~\cite{Cassel:2009pu}, the sign of the rotation angle $\zeta$ in Eq. (B-5) is opposite to the one used in the main text. 
In Ref.~\cite{Hook:2010tw}, where they cite Ref.~\cite{Cassel:2009pu}, 
the sign in the parenthesis in Eq. (9) should be opposite. 
} 
\beq
 m_{Z'_2}^2 = c_{\xi_1}^2 (\tilde{M}_V^2 )_{22} + s_{\xi_1}^2 m_Z^2 \lmk 1 - \frac{2 (\tilde{M}_V )^2_{12}}{t_{\xi_1} m_Z^2} \rmk \, . 
\eeq
In the main text of this paper, we simply denote the physical masses by $m_Z$, $m_{Z_{(B-L)_3}}$, and $m_{Z_h}$, noting $\xi_{1} \simeq s_{W} \epsilon_{1} / (1 - (m_{Z_{(B-L)_3}} / m_{Z})^{2}) \ll 1$ in most of the parameter space that we are interested in.

The current interactions are given by 
\beq
 {\cal L}_{\rm int} = (V^\mu)^{\rm T} \tilde{J}_\mu = (V^\mu_{\rm mass})^{\rm T} {\cal M} J_\mu\, , 
\eeq
where $V^\mu_{\rm mass} = (A^\mu, Z_1^\mu, Z_2^\mu, Z_3^\mu)^{\rm T}$, 
\beq
J^\mu \equiv 
\begin{pmatrix}
      e J_{\rm EM}^\mu \\ 
      g/ c_W J_{\rm Z^0}^\mu \\
      g_{(B-L)_3} J_{(B-L)_3}^\mu \\
      g_h J_{h}^\mu  
  \end{pmatrix}
  \equiv   \begin{pmatrix}
      c_W & s_W & 0\ & 0
      \\ -s_W & c_W & 0\  & 0 
      \\  0 & 0 & 1\ & 0 
      \\  0 & 0 & 0\ & 1       
    \end{pmatrix} 
    \tilde{J}^\mu\, , 
\eeq
and 
\beq
 {\cal M} &=& 
R_{31}^{\rm T}(\xi_3) 
R_{32}^{\rm T}(\xi_2) 
R_{12}^{\rm T}(\xi_1) 
   \begin{pmatrix}
      1 & 0 & 0 & 0 
      \\ 0 & 1 & 0 & 0 
      \\ - c_W t_{\epsilon_1}/c_{\epsilon_2} & s_W t_{\epsilon_1}/c_{\epsilon_2} & 1/(c_{\epsilon_1} c_{\epsilon_2}) & - t_{\epsilon_2} 
      \\ 0 & 0 & 0 & 1 
    \end{pmatrix} 
\nonumber\\
&\simeq& 
    \begin{pmatrix}
      1 & 0 & 0 & 0 
      \\ - c_W \epsilon_1 s_{\xi_1} & s_W \epsilon_1 s_{\xi_1} + c_{\xi_1} 
      & s_{\xi_1}  & - \epsilon_2 s_{\xi_1} - \xi_3
      \\ - c_W \epsilon_1 c_{\xi_1} \ & s_W \epsilon_1 c_{\xi_1}  - s_{\xi_1} \ 
      & c_{\xi_1} \ & - \epsilon_2 c_{\xi_1} - \xi_2 
      \\ - c_W \epsilon_1 (c_{\xi_1} \xi_2 + s_{\xi_1} \xi_3) \ & s_W  \epsilon_1 (c_{\xi_1} \xi_2 + s_{\xi_1} \xi_3) - s_{\xi_1} \xi_2 + c_{\xi_1} \xi_3 \ 
      &  c_{\xi_1} \xi_2 + s_{\xi_1} \xi_3 &  1 
    \end{pmatrix}\, . 
      \nonumber\\
\label{interactionM}
\eeq

Integrating out $Z_i$ ($i = 1,2,3$), 
one can check that 
the effective interactions between $J_h^\mu$ and $J_{\rm EM}^\mu$ or $J_{Z^0}^\mu$ 
are suppressed by a factor of order $g_h e \xi_1 \xi_2 / m_{Z_2}^2 \simeq g_h e  \epsilon_1 \epsilon_2 m_{Z_h}^2 / m_{Z_{(B-L)_3}}^4$. 
This implies that 
the leading-order interaction term, which is suppressed only by $g_h e  c_W \epsilon_1 \epsilon_2  / m_{Z_{(B-L)_3}}^2$, is cancelled in a large $m_{Z_3}$ limit. 
However, the momentum transfer for the process that we are interested in is of order $m_{Z_{3}}$. 
Thus we expect that 
the leading-order interaction is not completely cancelled 
out but is suppressed by a factor of $q^2 / (q^2 + m_{Z_3}^2)$, 
where $q^2$ is the square of the momentum transfer.

\section{Phase transition and thermal inflation}
\label{sec:appendix}

In this Appendix, we comment on the effects of the phase transition from the SSB of U(1)$\h$, in which the $U(1)_{h}$ gauge boson $\Psi$ develops the VEV $v_{\Psi}$.%
\footnote{
We do not discuss the phase transitions of the U(1)$_{(B-L)_i}$ breaking fields $\Phi_i$ 
because their potentials may be complicated by additional scalars and heavy fermions 
that are required to reproduce the proper Yukawa structure (see, e.g., Ref.~\cite{Alonso:2017uky}). 
}

Since it breaks the local Abelian gauge symmetry, 
cosmic strings form through the phase transition. 
However, 
their effects are negligible in our model 
because the energy density of cosmic strings is suppressed by a factor of the VEV squared in the Planck units 
when compared to the total energy density of the Universe.

If the VEV of a scalar field is much larger than its (zero temperature\,+\,thermal) mass, 
the potential energy before the phase transition may be much larger than the energy of the thermal plasma. 
In this case, 
the energy density of the Universe becomes dominated by the former vacuum energy 
and a mini inflation called a thermal inflation occurs through the phase transition~\cite{Yamamoto:1985rd, Lyth:1995ka}. 
The duration of the thermal inflation depends on the ratio between the VEV and (zero-temperature\,+\,thermal) mass of the SSB field. 
After the thermal inflation, 
the vacuum energy will be released into the radiation and the entropy production proceeds. 
As a result, the baryon asymmetry is diluted due to the entropy production at the time of this reheating. 

Here, we give a quantitative estimate of the dilution of the thermal relic through the entropy production. 
The (zero temperature\,+\,thermal) potential of $\Psi$ is given by 
\beq
 V(\Psi) = \lambda \lmk \abs{\Psi}^2 - \frac12 v_\psi^2 \rmk^2 + V_T(\Psi) \,, 
\eeq
where $\lambda$ is a quartic coupling. 
The mass of $\psi$ at the vacuum is given by $\sqrt{2 \lambda} \, v_\psi$. 
The thermal potential $V_T$ from $\psi$ and $Z\h$ is approximately given by 
\beq
 V_T (\Psi) = \lmk \frac{\lambda}{3} + q_\psi^2 \frac{g\h^2}{4} \rmk T^2 \abs{\Psi}^2 \,, 
\eeq
where $q_\psi$ ($=3$) is the charge of $\Psi$.  
At a high temperature, 
the thermal potential dominates and $\Psi$ is stabilized at $\Psi = 0$. 
When the temperature becomes lower than the critical temperature $T_c$, which is given by 
\beq
 T_c = \sqrt{\frac{\lambda }{\lambda/3 + q_\psi^2 g\h^2/4}} \, v_\psi \,, 
\eeq
the potential becomes unstable at $\Psi = 0$ 
and the scalar field starts to oscillate around the true vacuum at $\Psi = v_\psi / \sqrt{2}$. 
We define a dilution factor as the ratio of the initial to the final comoving 
entropy density as 
\beq
 \Delta \equiv \frac{s_f a_f^3}{s_i a_i^3} = 
 1 + \frac43 \frac{g_{* s} (T_{\rm RH, \psi})}{g_{*} (T_{\rm RH, \psi}) \, T_{\rm RH, \psi}} \frac{V (0)}{(2 \pi^2/45) \, g_{* s} (T_c) \, T_c^3 } \,, 
\eeq
where 
(I) $a_i$ ($a_f$) is the scale factor,
$s_i$ ($s_f$) is the entropy density before (after) the thermal inflation; (II)
$V(0) = \lambda v_\psi^4 / 4$; 
and (III) $g_{* s}$ ($g_{*}$) is the effective number of relativistic degrees of freedom for the entropy (energy) density as a function of $T$. 
The reheating temperature of oscillating $\psi$ satisfies $T_{\rm RH, \psi} < [30 V(0)/\{g_{*} (T_{\rm RH, \psi}) \, \pi^2\} ]^{1/4}$. 
To avoid the washout effect due to the thermal inflation, 
we require $\Delta \approx 1$ 
corresponding to
\beq
 \lmk \frac{m_{Z\h}^2}{m\h^2} + \frac{2}{3} \rmk \lesssim 
 \lambda^{-1/2} \lmk \frac{4 \sqrt{2} \pi^2 \, g_{* s} (T_c) \, g_{*} (T_{\rm RH, \psi})}{15 \, g_{* s} (T_{\rm RH, \psi})} \rmk^{2/3} \lmk \frac{15}{2 \pi^2 g_{*} (T_{\rm RH, \psi})} \rmk^{1/6} \,. 
\eeq
It follows that the gauge boson mass cannot be arbitrary larger than the mass of the SSB field. 
This condition is easily satisfied in our model although it is non-trivial in other models 
with hierarchical mass scales.

\bibliography{reference}

\end{document}